# A Survey of Different Residential Energy Consumption Controlling Techniques for Autonomous DSM in Future Smart Grid Communications

M. N. Ullah[1], A. Mahmood[1], S. Razzaq[2], M. Ilahi[1], R. D. Khan[3], N. Javaid[1]

[1]COMSATS Institute of Information Technology, Islamabad, Pakistan.
[2]COMSATS Institute of Information Technology, Abbottabad, Pakistan.
[3]COMSATS Institute of Information Technology, Wah Cant, Pakistan.

**ABSTRACT**
In this work, we present a survey of residential load controlling techniques to implement demand side management in future smart grid. Power generation sector facing important challenges both in quality and quantity to meet the increasing requirements of consumers. Energy efficiency, reliability, economics and integration of new energy resources are important issues to enhance the stability of power system infrastructure. Optimal energy consumption scheduling minimizes the energy consumption cost and reduce the peak-to-average ratio (PAR) as well as peak load demand in peak hours. In this work, we discuss different energy consumption scheduling schemes that schedule the household appliances in real-time to achieve minimum energy consumption cost and reduce peak load curve in peak hours to shape the peak load demand.

## I. INTRODUCTION

A system that implements communication and information technology in electrical grid is known as the smart grid. Smart grid system gathers information about the activities of electrical energy supplier and user. In order to improve the reliability, economics and efficiency of the electrical energy production and distribution, smart grid takes action in an autonomous manner.

Smart grid improves the customer's load utilization by deploying the demand side management (DSM) programs. DSM monitors, plan and implement those utility functionalities that are designed to incentivise the consumers during utilization of electricity in respect to shape the utility load curve. Demand side management programs are implemented to exploit the better utilization of current available generating power capacity without installing the new power generation infrastructure [1]. DSM was proposed in the late 1970s [2]. Worldwide energy consumption in buildings is approximately 40% of global power consumption. Better utilization of energy consumption in buildings is an important issue. DSM programs enable two way communication b/w users and grid to schedule household appliances and co-ordination. Heavy loads run in the peak hours due to which utility load curve is potentially going high. Therefore, DSM can control the residential loads by shifting the load from peak hours to off-peak hours in order to reduce the peak load curve and improve energy efficiency by scheduling the energy consumption.

In previous literature, various load management techniques have been discussed to enable autonomous DSM in future smart grid. An incentive based energy consumption controlling scheme discussed in [3]. A direct load control (DLC) scheme for residential load control discussed in [4] & [5]. In [6], a model is proposed to schedule the household appliances.

*****Corresponding Author:** Nadeem Javaid, COMSATS Institute of IT, Islamabad, www.njavaid.com.

## II. PEAK LOAD REDUCTION

Energy efficiency refers to using less energy to provide the same or an improved level of service to the energy consumer in an economically efficient way, it includes using less energy at any time, including during peak periods. To achieve high reliability in electric grid utility companies needs to reduce the peak load demand. Now a day's customer's expectations are increasing both in quality and quantity. Limited energy assets and expensive process of integrating new energy resources, there is an important need to improve our power system utilization. By increasing the different types of new loads e.g. plug-in hybrid electric vehicles (PHEVs) which can potentially increase the normal residential load. By this, it is very important to develop new methods for peak load reduction. There are some environmental issues related to current power systems. Some countries widely use the oil and coal fired power plants to meet the peak demands, a huge amount of $CO_2$ and greenhouse gases are emitted. It could potentially be minimized by using by using efficient demand side management (DSM) program. DSM has been introduced since the late 1970s. It can be used for peak shaving to load. In the following subsections we discuss different techniques for peak load reduction.

Previous literature employs different techniques for peak load reduction in smart grid.

### II.1 AN AUTONOMOUS THREE LAYERED STRUCTURAL MODEL FOR DEMAND SIDE MANAGEMENT

In [2], authors discuss a scheduling model which consisting of three modules i.e admission controller, load balancer and demand response manager that is used to control the peak load demand. The model of this scheme is shown in Fig. 1 in [2].

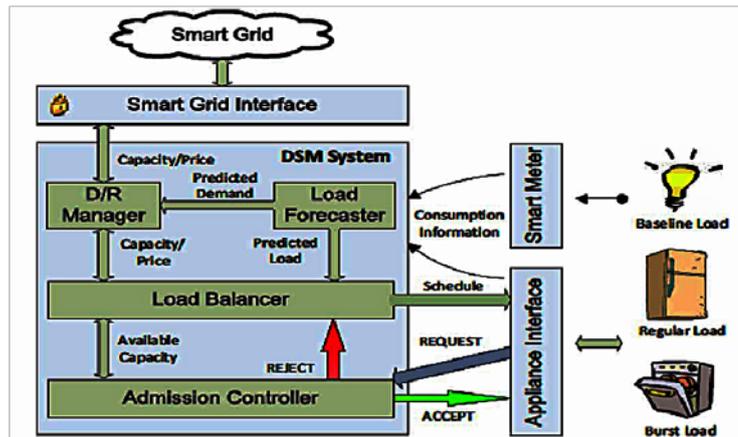

Fig. 1. The proposed model for load management on demand side

Powers loads are divided into three different categories based on their load characteristics for the minimization of peak load demand.

   I.   Baseline load is the power consumption of those appliances that must be run immediately at any time. Baseline load includes lighting, cooking rings and networking devices.
   II.  Burst load is the power consumption of those appliances that ON for a fixed duration and start and stop within the given deadline e.g., washing machine, dishwasher and dryer.
   III. Regular load is related to appliances that are always in running position during a long period, such as refrigerator, water heater and HVAC (Heat Ventilation Air Conditioning) of the house.

Run-time scheduling is used to control the appliances operation in order to limit the power capacity. The spring algorithm is used for scheduling the devices on run time. For the scheduling of appliances, the spring algorithm is determined by the heuristic function H, called heuristic value. Heuristic value is the balanced vector between 0 and 1, which represent the essentiality of task. At each level of searching the time slot, the function H is applied to each remaining task residual to be scheduled. In this algorithm we consider the case where task priority may change during its execution time. Simulation results shown in Fig.2 and 3 [2].

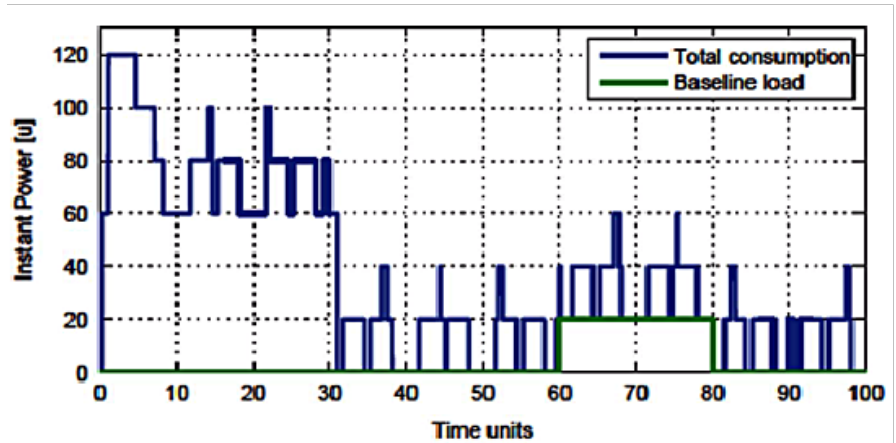

Fig. 2. Energy consumption without load management (Baseline load and total consumption)

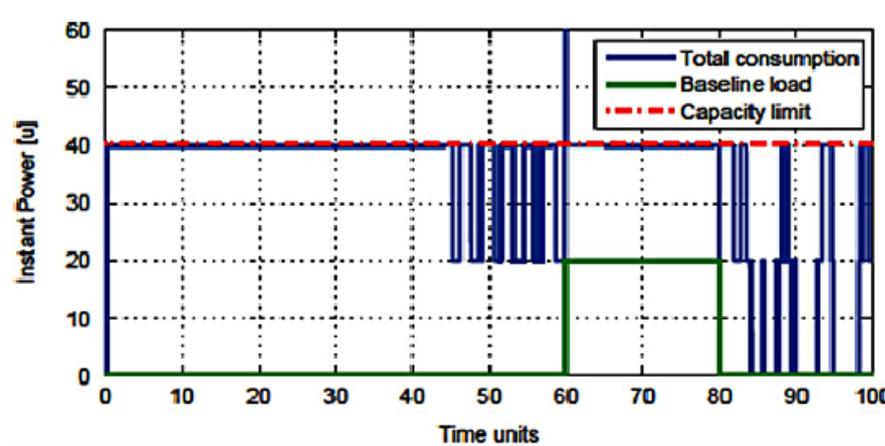

Fig. 3. Peak load reduction by online scheduling (Total energy consumption and baseline load)

### II.2        BACKTRACKING-BASED TECHNIQUE FOR LOAD CONTROL

A task model is presented in [7], to schedule the home appliances for reducing the peak load as well as the global peak reduction to enable the demand side management. This model consists of actuation time, operation length, deadline, and consumption profile. Scheduler copies the profile entry of different appliances one by one according to task profile of the allocation table. This model reduces the peak load up to 23.1 %. The task model shown in Fig.4 [7].

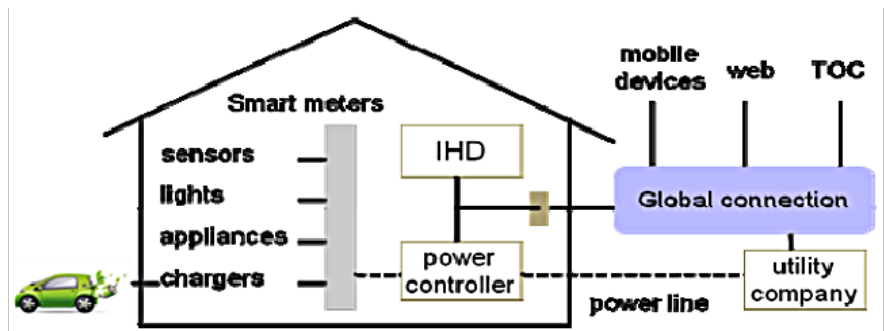

Fig. 4. Power scheduler operation

The backtracking optimization technique is used for scheduling the appliances. Backtracking additionally frame a search tree. This potential search tree consists of all feasible solutions including worthless solutions. At each intervening node, which passes to a feasible solution, it checks either the node can guide to a feasible solution. If it cannot, remaining sub tree is reduced. Otherwise iteration proceeds to the next level. By this phenomena scheduler search the feasible time slots for the appliances schedule. In this way, peak load reduce as well as global peak.

### II.3 GAME-THEORETIC BASED DEMAND SIDE MANAGEMENT

A game theory technique in [8], proposes an energy consumption scheduling game. In this game-theoretic model users are the players and their daily schedule of using appliances and loads are strategies. Incentives are provided to users to participate in this energy consumption scheduling game. Block diagram of this model is shown in Fig.5 [8].

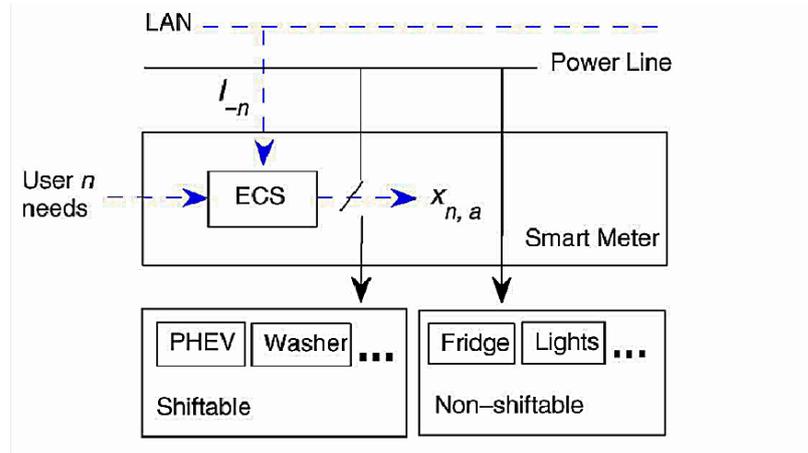

Fig. 5. Model with ECS devises deployment

ECS devices are used in smart meters to schedule the appliances. By deployment of ECS function, load is equally distributed across different hours of the whole day. By minimizing the energy cost of all users, we also minimize the peak load. The optimal solution can be achieved by solving the following optimization problem:

$$\underset{x_n \in X_n, \forall n \in N}{\text{minimize}} \sum_{h=1}^{H} C_h (\sum_{n \in N} \sum_{a \in A_n} x_{n,a}^h) \tag{1}$$

Where

$C_h(\cdot)$    Cost function

$x_{n,a}^h$    Energy consumption of appliance 'a' in hour $h$.

This optimization problem solved by integer programming method (IPM) to find the optimal solution. Round-robin scenario is used in ECS devices algorithm. Simulation results show that with the deployment of ECS function of smart meters, peak-to-average ratio reduce up to 17%.

### II.4 ECS BASED SCHEDULING ALGORITHM

In [9], an ECS based scheduling technique is discussed. In this technique, consider a scenario of power system where an energy source (e.g., a generator is connected to electric grid) is shared by different users each one deployed ECS device in the smart meter shown in fig 6 [9].

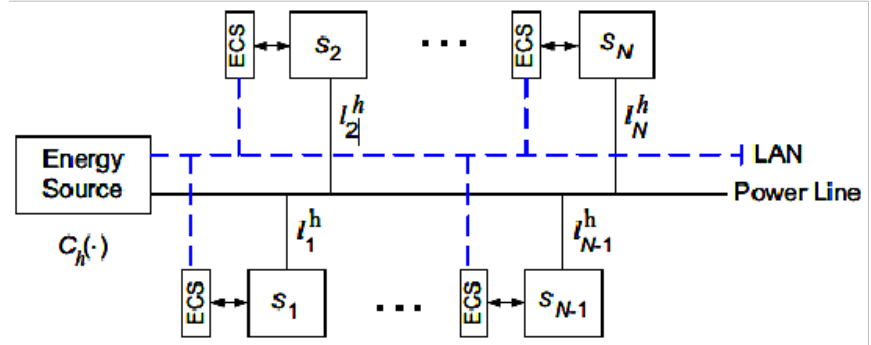

Fig. 6. Smart grid system model with N load subscribers

Assume that each user is facilitated with a smart meter. Energy consumption scheduling devices (ECS) are deployed in smart meters to meet the peak load reduction. ECS devices are interrelating automatically by running an algorithm to find the optimal schedule for energy consumption of each subscriber. In this way, reduce peak load demand.

### II.5 AN OPTIMAL AND AUTONOMOUS RESIDENTIAL LOAD CONTROL SCHEME

In [1], an automatic and optimal energy consumption scheduling scheme is discussed to minimize the peak-to-average ratio in residential areas. In this scenario, real-time pricing and inclining block rates are combined to balance the load and minimize peak-to-average ratio. An energy consumption scheduler is deployed in residential smart meters to control the load of household appliances. Price predictor is used in smart meters to estimate upcoming prices rates. When load demand is high in smart grid, requests sent by grid to smart meters to reduce the load. In this scenario, scheduler takes action and increases the upcoming prices of next 2 or 3 hours by optimization technique. Therefore, some portion of load automatically suspends and reduces the total load.

### III. MONETARY COST MINIMIZATION

In the future smart grid both power companies and users can take advantage of economical and environmental aspects of smart pricing models [1]. Efficiency of electricity power consumption is an important factor. Consumer's energy demand increase during peak hours. Utility companies run peaker power plants to meet this peak demand during peak hours. Peaker plants charge higher prices per kilowatt hour. Communication and metering technologies enable smart devices in homes to reduce peak load demand during high cost peak hours. Electricity prices increased during peak hours and low during off peak hours. Therefore consumers avoid high price peak hours and shift their heavy loads to off peak hours. Currently, consumption of electricity in buildings is not in efficient way that leading towards the wastage of billions of dollars and huge amounts of greenhouse gas emission. General approach for the management of energy consumption in buildings: shifting consumption and reducing consumption. Smart grid applies demand side management (DSM) to cope with the energy consumption cost. An increasing number of (Phevs) leading to the unbalanced conditions in the utility load shape, voltage problem and cost maximization. Energy consumption cost can be minimized by shifting the high load household from peak hours to off peak hours. In previous literature, following techniques are discussed to minimize the monetary cost.

### III.1 A THREE LAYERED STRUCTURE MODEL FOR AUTONOMOUS DEMAND SIDE MANAGEMENT

In [2], the proposed scheme adopts a layered structure shown in Fig.1 [2] which enables load management in smart buildings. Power load request scheduling is composed as mixed integer programming (MIP) problem, which reduces the overall operational cost. Admission controller using optimization technique to minimize the operational cost in respect to capacity constraints specify by the demand response manager (DRM) & operational constraints defined for each appliance. Careful energy management of burst loads is an important issue for the energy consumption cost on demand side. Load balancer (LB) evenly distributes the electrical load of appliances over a time frame and schedules the requests that have been refused by admission controller. Therefore load balancer minimizes the cost function analogous to energy price. Each time load balancer is activated. To minimize the total energy cost, a mixed integer programming (MIP) problem is solved. In different circumstances of considered appliances, load balancer establishes an appropriate schedule that would evenly distribute the appliances load over a time horizon. In such a way overall energy cost is minimized under different capacity constraints.

### III.2 BACKTRACKING-BASED TECHNIQUE FOR LOAD CONTROL

The proposed task model in [7], schedule the home appliances in way to reduce peak load demand as well as energy cost. Present scheme design a power scheduler capable of minimizing peak load in household consumption or buildings. This model consists of actuation time, operation length, dead line and consumption profile. Scheduler copies the profile entry of different appliances one by one according to task profile to the allocation table. Controller exchanges the price information and load demand with utility company. Information exchange via cellular network or internet. In response the utility company sends information about residential load change enabling demand response, load control and price adjustment. Appliances are characterizes into 3 classes on the basis of their power consumption profile. Class 1 tasks should be start once after the task gets ready & cannot terminated to be end e.g. hair dryer. Class 2 tasks gets ready but not start at that moment and they have non-preemptive operation e.g. dish washer, dryer, laundry machine. Class 3 tasks start after their activation time but they have preemptive operation e.g. PHEVs charging belong to this category. This scheme only schedule preemptive and non-preemptive tasks. This model used Backtracking-based scheduling strategy to find an optimal scheduling in homes. The appliances (tasks) to be scheduled are less than 10. Backtracking-based scheme schedule the tasks on different time slots in a given time domain. Optimal scheduling of appliances reduces the peak load curve and also reduces the energy consumption cost.

### III.3 GAME-THEORETIC BASED DEMAND SIDE MANAGEMENT

Game-theoretic based model in [8], develop an energy consumption game for cost minimization. In this model users are players and their strategies are their daily scheduling loads. It also considered that utility company can adopt suitable pricing tariffs that characterize the energy usage both in time and level. Optimal performance in terms of energy cost minimization is achieved at Nash equilibrium of energy scheduling game. We consider a common scenario, single utility company serves different users. This model systematically manages the appliance schedule and shifts them in order to reduce energy cost. For an efficient energy consumption scheduling, energy cost minimization to all users express in the following optimization problem.

### III.4 ECS BASED SCHEDULING ALGORITHM

In [9], an ECS based residential energy consumption controlling scheme is discussed. In this scheme, deployment of energy consumption scheduling (ECS) devices in smart meters enable autonomous demand side management (DSM). Distributed algorithm is used to schedule the optimal energy consumption for each subscriber. This algorithm reduces the total energy cost. A new pricing scheme, develop from game-theoretic model to reduce the total cost of the electric power system. A model of a smart grid system is shown in figure. Subscribers are denoted by N. Simulation results show that the deployment of ECS devices reduces the energy consumption cost. Simulation results shown in Fig.7 and [9].

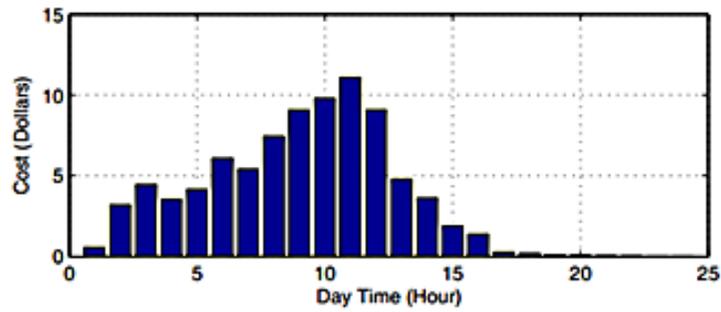
Fig. 7. Daily cost $ 86.47 (ECS devices are not used)

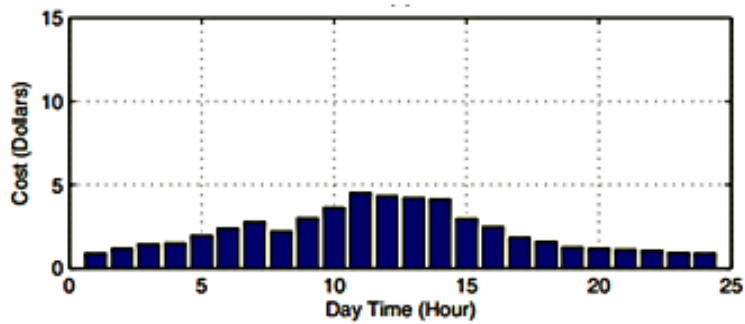
Fig. 8. Daily cost $ 53.81 (ECS devices are deployed)

### III.5 VICKREY-CLARKE-GROVES (VCG) MECHANISM BASED DSM

Vickrey-Clarke-Groves (VCG) mechanism in [10], maximizes the social welfare i.e. the difference between aggregate utility function of all users and total energy cost. We consider that each user deployed energy consumption controller (ECC) device in its smart meter. VCG mechanism develops the DSM programs to enable efficient energy consumption among all users. In this scheme each user provides its energy demand to the utility. By deploying a centralized mechanism in ECC device, the energy provider estimates each user's optimal energy consumption level and declares particular electricity payment for each user. In this way VCG mechanism reduces the energy cost. For optimal solution we evolve an optimization problem to reduce the total energy cost charged on energy provider while maximizing aggregate utility functions of all users.

### III.6 A SCHEME FOR TACKLING LOAD UNCERTAINTY

In [11], an optimization based residential load controlling algorithm is proposed that tackling the load irregularity to reduce energy cost in real-time. Some analytical assessment is required about future load demand in this model. In this scenario both real-time and inclining block rate prices strategies are combined. The proposed algorithm is formulated as an optimization problem. Energy cost can be minimized by solving this optimization problem. Each appliance sends request to control unit that schedule energy consumption requests in different time slots. The appliance starting time of operation depends on the request acceptance. The control unit determine a suitable schedule of each appliance in each time slot of giving time domain.

### IV. CONCLUSION

In this paper, we discussed different residential load controlling techniques in the smart grid. These techniques reduce the energy consumption and also minimize peak-to-average ratio (PAR) as well as peak load by shifting the heavy loads from peak-hours to off peak-hours. Consumer should also be incentivize to schedule the appliances according to schemes discussed in the paper. Scheme 1 reduce the peak load up to 66.66%. So this model is more efficient. ECS based and Vickrey-Clarke Groves mechnism minimize the cost up to 37%.